\documentstyle{mn}

\newif\ifAMStwofonts

\font\caps=cmcsc10  

\def\be{\begin{equation}}
\def\ee{\end{equation}}
\def\mathnew{\mathsurround=0pt}
\def\simov#1#2{\lower .5pt\vbox{\baselineskip0pt \lineskip-.5pt
        \ialign{$\mathnew#1\hfil##\hfil$\crcr#2\crcr\sim\crcr}}}
\def\simgreat{\mathrel{\mathpalette\simov >}}
\def\simless{\mathrel{\mathpalette\simov <}}
\def\kms{{\rm\,km\,s^{-1}}}
\def\kpc{{\rm\,kpc}}
\def\msun{{\rm\,M_\odot}}

\def\yr{{\rm\,yr}}

\def\gmcs{{\caps gmc}s}

\ifoldfss
  \ifCUPmtlplainloaded \else
    \NewTextAlphabet{textbfit} {cmbxti10} {}
    \NewTextAlphabet{textbfss} {cmssbx10} {}
    \NewMathAlphabet{mathbfit} {cmbxti10} {} 
    \NewMathAlphabet{mathbfss} {cmssbx10} {} 
  \fi
  \ifAMStwofonts
    \ifCUPmtlplainloaded \else
      \NewSymbolFont{upmath} {eurm10}
      \NewSymbolFont{AMSa} {msam10}
      \NewMathSymbol{\upi}     {0}{upmath}{19}
      \NewMathSymbol{\umu}     {0}{upmath}{16}
      \NewMathSymbol{\upartial}{0}{upmath}{40}
      \NewMathSymbol{\leqslant}{3}{AMSa}{36}
      \NewMathSymbol{\geqslant}{3}{AMSa}{3E}

    \fi
  \fi
\fi 

\ifnfssone
  \newmathalphabet{\mathit}
  \addtoversion{normal}{\mathit}{cmr}{m}{it}
  \addtoversion{bold}{\mathit}{cmr}{bx}{it}
  \newmathalphabet{\mathbfit} 
  \addtoversion{normal}{\mathbfit}{cmr}{bx}{it}
  \addtoversion{bold}{\mathbfit}{cmr}{bx}{it}
  \newmathalphabet{\mathbfss} 
  \addtoversion{normal}{\mathbfss}{cmss}{bx}{n}
  \addtoversion{bold}{\mathbfss}{cmss}{bx}{n}
  \ifAMStwofonts
    \ifCUPmtlplainloaded \else
      \UseAMStwoboldmath
      \makeatletter
      \new@mathgroup\upmath@group
      \define@mathgroup\mv@normal\upmath@group{eur}{m}{n}
      \define@mathgroup\mv@bold\upmath@group{eur}{b}{n}
      \edef\UPM{\hexnumber\upmath@group}
      \new@mathgroup\amsa@group
      \define@mathgroup\mv@normal\amsa@group{msa}{m}{n}
      \define@mathgroup\mv@bold\amsa@group{msa}{m}{n}
      \edef\AMSa{\hexnumber\amsa@group}
      \makeatother
      \mathchardef\upi="0\UPM19
      \mathchardef\umu="0\UPM16
      \mathchardef\upartial="0\UPM40
      \mathchardef\leqslant="3\AMSa36
      \mathchardef\geqslant="3\AMSa3E
    \fi
  \fi
\fi 

\ifnfsstwo
  \DeclareMathAlphabet{\mathbfit}{OT1}{cmr}{bx}{it}
  \SetMathAlphabet\mathbfit{bold}{OT1}{cmr}{bx}{it}
  \DeclareMathAlphabet{\mathbfss}{OT1}{cmss}{bx}{n}
  \SetMathAlphabet\mathbfss{bold}{OT1}{cmss}{bx}{n}
  \ifAMStwofonts
    \ifCUPmtlplainloaded \else
      \DeclareSymbolFont{UPM}{U}{eur}{m}{n}
      \SetSymbolFont{UPM}{bold}{U}{eur}{b}{n}
      \DeclareSymbolFont{AMSa}{U}{msa}{m}{n}
      \DeclareMathSymbol{\upi}{0}{UPM}{"19}
      \DeclareMathSymbol{\umu}{0}{UPM}{"16}
      \DeclareMathSymbol{\upartial}{0}{UPM}{"40}
      \DeclareMathSymbol{\leqslant}{3}{AMSa}{"36}
      \DeclareMathSymbol{\geqslant}{3}{AMSa}{"3E}
    \fi
  \fi
\fi 

\ifCUPmtlplainloaded \else
  \ifAMStwofonts \else 
    \def\upi{\pi}
    \def\umu{\mu}
    \def\upartial{\partial}
  \fi
\fi

\title{Relaxation in stellar systems and the shape and rotation of the inner
dark halo}

\author[S. Tremaine and J. P. Ostriker]
{Scott Tremaine and Jeremiah P. Ostriker \\
Princeton University Observatory, Peyton Hall, \\
Princeton, NJ~08544-1001, USA}

\pagerange{\pageref{firstpage}--\pageref{lastpage}}

\begin{document}

\maketitle
\label{firstpage}

\begin{abstract}

\noindent
Why do galactic bars rotate with high pattern speeds, when dynamical friction
should rapidly couple the bar to the massive, slowly rotating dark halo? This
longstanding paradox may be resolved by considering the dynamical interactions
between the galactic disc and structures in the dark halo. Dynamical friction
between small-scale halo structure and the disc spins up and flattens the
inner halo, thereby quenching the dynamical friction exerted by the halo on
the bar; at the same time the halo heats and thickens the disc, perhaps
forming a rapidly rotating bulge. Two possible candidates for the required
halo structures are massive black holes and tidal streamers from disrupted
precursor halos. More generally, gravitational scattering from phase-wrapped
inhomogeneities represents a novel relaxation process in stellar systems,
intermediate between violent relaxation and two-body relaxation, which can
isotropize the distribution function at radii where two-body relaxation is not
effective.

\end{abstract}


\section{Introduction}

\label{sec:intro}

Most of the mass in disc galaxies resides in three components: a thin, rapidly
rotating disc of stars and gas, a centrally concentrated stellar bulge or
spheroid, and an invisible dark halo that extends far beyond the outer edge of
the stellar disc. The bulge rotates less rapidly than the disc, and the
rotational properties of the halo are unknown. The halo is believed to form by
hierarchical clustering of dark matter, which results from gravitational
instability of small irregularities present at much earlier times; its
structure depends on the cosmological model, but usually the halo is strongly
triaxial \cite{DC91,WQSZ92} and rotates only slowly (median spin parameter
$\lambda\sim 0.05$; see for example Steinmetz \& Bartelmann \nocite{SB95}
(1995).  The disc is believed to form from baryonic material that collects in
the potential well of the dark halo. Numerical simulations
\cite{KG91,D94,KB98} show that the growth of the disc modifies the structure
of the inner halo (by which we mean the part at radii smaller than the outer
edge of the disc): the halo becomes more nearly axisymmetric ($b/a\simgreat
0.8$) and oblate ($c/a \simeq 0.6$) but remains slowly rotating.

In this paper we examine the long-term gravitational interactions between the
disc and inner halo. In particular we shall discuss the rate and consequences
of angular momentum transfer from the disc to the halo, which tends to flatten
and spin up the inner halo, while causing the disc material to thicken, heat,
and drift towards the centre. Any irregularities in either the disc or halo
tend to contribute to angular momentum transfer; we shall argue that the
transfer is probably sufficiently strong to transform the inner halo from a
slowly rotating oblate spheroid supported by pressure anisotropy into a
rapidly rotating thickened disc.

Rapid rotation of the inner halo may resolve a persistent puzzle about the
structure of barred disc galaxies. A rotating bar embedded in a non-rotating
dark halo loses angular momentum to the halo through dynamical friction, on a
time-scale that is generally much less than a Hubble time.  Thus bars should
have small pattern speeds, corresponding to large values of the dimensionless
ratio ${\cal R}\equiv$(corotation radius)/(radial extent of the bar)--which
is normally $\simeq 1$ when a bar forms. This argument was first explicitly
discussed by Weinberg (1985)\nocite{W85}, although earlier numerical models of
disc galaxies with live halos by Sellwood (1980)\nocite{S80} exhibited rapid
angular momentum transfer from the disc to the halo after a bar formed in the
disc. Hernquist \& Weinberg (1992)\nocite{HW92} simulated systems containing
a non-rotating halo and a rigid bar (with no disc) and concluded that the
gravitational coupling was `sufficiently strong to remove all of the bar's
angular momentum on a time-scale much shorter than a Hubble time'. Little \&
Carlberg (1991a,b) \nocite{LC91a,LC91b} followed the long-term evolution of
barred disc galaxies with live halos using a two-dimensional code, which
greatly improves spatial resolution but requires a somewhat unrealistic flat
halo. They found that ${\cal R}$ increased from 1 to 2 over $\sim 30$ initial
bar rotation periods. The most recent and comprehensive simulations of this
process are by Debattista \& Sellwood (1998),\nocite{DS98} who followed
self-consistent $N$-body models of a barred disc and initially spherical halo.
In the simulation with the most massive halo, they found that the bar pattern
speed dropped by a factor of five (to ${\cal R}\simeq 2.3$) within 50 initial
bar rotation periods. Only in models with low-density halos did the bar pattern
speed remain high.

Thus a wide range of analytic arguments and numerical simulations concur that
high-density halos strongly decelerate bars. On the other hand, observations
show that most bars are rapidly rotating. Bar pattern speeds can be estimated
in several ways: (i) By comparing the shocks in models of gas flow with dust
lanes in barred spiral galaxies. Using this approach Athanassoula (1992)
\nocite{A92} estimated ${\cal R}=1.2\pm0.2$. (ii) From the rotation field of
the old disc population and the equation of continuity. This method has been
applied to the SB0 galaxy NGC 936 and yields ${\cal R}\simeq1.4\pm0.3$
\cite{KG89,MK95}. (iii) By identifying photometric and kinematic properties of
stars and gas with Lindblad or corotation resonances; this approach is
probably less reliable since some disc features may be transitory and also the
pattern speed of the spiral structure may differ from that of the bar
\cite{SS88}. A recent review by Elmegreen (1996)\nocite{E96} concludes that
${\cal R}=1.2\pm0.2$ in early-type galaxies; bars in late-type galaxies may
rotate more slowly but so far there are very few observations.

Thus, there is a significant and growing contradiction between theory (${\cal
R} \gg 1$) and observations (${\cal R}\sim 1$). Debattista \& Sellwood (1998)
\nocite{DS98} argue that the absence of slowly rotating bars can only be
explained if the halo contributes less than $\sim 25\%$ of the radial force in
the region where bars are found (around 2 disc scale lengths); this in turn
requires that the disc provides almost 90\% of the total circular speed
($\gamma\equiv v_{\rm disc}/v_{\rm tot}>0.87$; see \nocite{O95} Olling 1995
for a more precise definition). The relative contribution of the disc and dark
halo to the radial force in spiral galaxies is controversial and uncertain
\cite{B97,BC97,CR97,S97,F98,S98}, but it seems unlikely that the very high
value of $\gamma$ required by Debattista and Sellwood is present in most disc
galaxies. For example, a careful recent discussion of mass models for the
Galaxy is given by \nocite{OM98} Olling \& Merrifield (1998); the nine models
they describe with various values for the solar circular speed and radius have
$\gamma$ in the range 0.66--0.83 and their favored model has $\gamma=0.72$.
Even `maximal disc' models of external galaxies--which have the largest
possible disc mass consistent with the observed rotation curve and an
isothermal dark halo--have a median $\gamma$ of only 0.85 \cite{S97}, which is
slightly lower than Debattista and Sellwood require.

A rapidly rotating inner halo could resolve this apparent contradiction, by
reducing or even reversing the drag on a rotating bar\footnote{Debattista \&
Sellwood (1998) \nocite{DS98} find that halo rotation is not sufficient to
suppress bar slowdown, but their rotating halos are constructed by reversing
the angular momenta of stars in a spherical halo, and thus do not rotate as
rapidly as a flattened halo.}.

\section{Disc-halo gravitational torques}

Disc-halo angular momentum transfer can result from structure in either the
disc or halo. We first examine the effects of disc structure (\S
\ref{sec:disc}) and find that no known disc features transfer angular momentum
rapidly enough to spin up the inner halo. Halo structures
(\S \ref{sec:halo}) are more efficient. 

\subsection{Structure in the disc}

\label{sec:disc}

The discs of spiral galaxies exhibit a wide variety of structure, including
molecular clouds, spiral arms, and central bars. All of these can transfer
angular momentum to the halo.

To estimate the transfer rate we adopt a simple model in
which the circular speed of the disc, $v_c$, is independent of radius, and a
fraction $f_h$ of the gravitational force in the disc plane is due to a
spherical halo; the remaining fraction $f_d=1-f_h$ arises from the self-gravity
of the disc. In this model the halo density and disc surface density are
\cite{BT87}
\be
\rho_h(r)=f_h{v_c^2\over 4\pi Gr^2}, \qquad \Sigma_d(r)=f_d {v_c^2\over 2\pi G
r}.
\label{eq:isodef}
\ee
If the torque per unit disc area exerted on the halo is $N$, then the
disc material in the radius range $[r,r+dr]$ loses angular momentum at the
rate $2\pi r N dr$; if we assume that this angular momentum is gained by the
halo material in the same radius range, with mass $4\pi r^2 \rho_h(r) dr$ and
characteristic specific angular momentum $\sim rv_c$, then the halo is 
expected to flatten on a characteristic time
\be
\tau \equiv  {2\rho_h(r) r^2v_c\over N}={f_hv_c^3\over 2\pi GN}.
\label{eq:taudef}
\ee

First consider the effects of discrete masses in the disc. We suppose that a
fraction $f_M$ of the disc mass is in objects of mass $M$ and radius $R$. Each
such object loses specific angular momentum to the halo by dynamical friction,
at a rate (Binney \& Tremaine 1987, \nocite{BT87} eq. 7-24)
\be
\dot L= -0.43f_h{GM\over r}\ln\Lambda,
\ee
where $\Lambda\approx r/R$. The torque per unit area on the halo is then 
$N=f_M\Sigma_d|\dot L|$, and the halo flattening time is 
\begin{eqnarray}
\tau & = &  2.3{v_cr^2\over f_Mf_dGM\ln\Lambda}\nonumber \\
& = & {1\times10^{12}\yr}\,\left(0.1\over f_Mf_d\right)
\left(v_c\over 200\kms\right)\left(r\over
3\kpc\right)^2 \nonumber \\
& & \qquad \times \left(10^6\msun\over M\right)\left(10\over\ln\Lambda\right).
\label{eq:gmc}
\end{eqnarray}
The largest disc objects are giant molecular clouds (\gmcs). These have a wide
range of masses but their contribution to dynamical friction depends on the
weighted average $\langle M^2\rangle/\langle M \rangle$, which is dominated by
the largest clouds at mass roughly $10^6 \msun$ (e.g. \nocite{C91} Combes
1991). The largest \gmcs\ may not be bound, but unbound objects transfer
angular momentum just as well. The fraction of the total mass in molecular gas
ranges from 4\% for Sa galaxies to 25\% for Scd galaxies \cite{YS91} so
$f_Mf_d=0.1$ is a reasonable average. We then see from equation (\ref{eq:gmc})
that \gmcs\ do not transfer significant angular momentum to the bulk of the
inner halo (although individual \gmcs\ at radii $\simless 1\kpc$ do spiral
inwards in less than a Hubble time; see \nocite{SGBB91} Stark et al. 1991).

Next we examine angular momentum transfer by spiral structure in the disc.  A
tightly wrapped spiral pattern with azimuthal wavenumber $m$, pattern speed
$\Omega_p$, radial wavenumber $k$ and fractional surface-density amplitude
$\Sigma_1/\Sigma_d$ exerts a torque per unit area on the halo
\cite{M76}
\be
N={2^{7/2}\pi^{3/2}\over 3}{m^2\Omega_p\rho_hG^2\Sigma_d^2\over k^4\sigma^3}
\left(\Sigma_1/\Sigma_d\right)^2,
\ee
where $\sigma$ is the one-dimensional velocity dispersion in the halo and the
halo distribution function is assumed to be Maxwellian. In our model
$\sigma=v_c/2^{1/2}$, and using equations (\ref{eq:isodef}) and
(\ref{eq:taudef}) the halo flattening time is
\be
\tau={3\pi^{1/2}\over 4} {m^2\over \Omega_p f_d^2}\left(\cot i\right)^4
\left(\Sigma_d\over\Sigma_1\right)^2; 
\ee
where $i=\tan^{-1}(m/kr)$ is the pitch angle of the pattern. 
An optimistic set of parameters is $m=2$, $i=20^\circ$, $f_d\simeq 0.7$,
$\Omega_p^{-1}\simeq 5\times 10^7$ yr, and $\Sigma_1/\Sigma_d\simeq 0.2$,
which yields $\tau\simeq 10^{12}$ yr, too long to be of interest. Unusually
strong and open spiral patterns could transfer angular momentum more rapidly,
but these are normally transitory, induced for example by an encounter with 
a nearby galaxy. 

Angular momentum transfer by dynamical friction on a central bar has been
discussed in the Introduction. Although the frictional forces are strong
enough to despin the bar, it is much harder to spin up the halo, because the
reservoir of angular momentum in the bar is too small. Bars typically contain
$\simless 30\%$ of the disc luminosity and extend only to $\sim 30\%$ of the
disc radius, so their moment of inertia is only a few percent of the disc's;
thus transferring all of the bar's angular momentum to the halo--as noted, a
rapid process--would still impart negligible halo rotation.

We conclude that the known disc structures cannot spin up the
inner halo. 

\subsection{Structure in the halo}

\label{sec:halo}

The dark halo is believed to form by gravitational instability of small
density fluctuations in the early Universe. In standard hierarchical models,
dense, low-mass halos form first, and then merge to build up successively more
massive but less dense objects. The process terminates at a redshift $z$ given
roughly by $1+z\simeq \Omega_m^{-1}$, where $\Omega_m$ is the density
parameter. When a small halo merges with a larger one, it loses energy by
dynamical friction and spirals towards the centre until it is tidally stripped
or completely disrupted, at a radius $r\sim r_s(M/M_s)^{1/3}$, where $r_s$ and
$M_s$ are the radius and mass of the small halo and $M$ is the mass of the
larger halo interior to $r$. The disrupted halo is stretched into a tidal
streamer; the length of the streamer after a time $t$ is roughly $\Delta\phi
\sim (t/P)(r_s/r)\sim (t/P)(M_s/M)^{1/3}$, where $P$ is the orbital period
\cite{T93,J98}. Such streamers can survive as distinct--though
unbound--structures in phase space for a Hubble time or longer; since the
halo material is collisionless the streamers never overlap in phase space but
simply become longer as phase mixing proceeds, forming a kind of phase-space
spaghetti. Structure resulting from incomplete phase-mixing is already
well-known in several other contexts: moving groups in local disc stars
\cite{E65,D98}, shells in elliptical galaxies \cite{HQ88,Mal98}, clumps in
kinematic surveys of the metal-weak halo \cite{MMH96}, and the disrupted
Sagittarius dwarf galaxy \cite{I97}.

A rotating disc exerts dynamical friction on a tidal streamer. To estimate the
frictional force, consider the idealized case in which a single streamer of
length $L\simless r$ pierces the disc at right angles. At distances $\ll L$,
we can approximate the streamer as an infinite, straight wire, and we can
approximate the wire as stationary since the velocity vector of its material
is approximately aligned with the streamer. We can approximate the disk as a
uniform, infinite, zero-thickness sheet moving past the streamer at velocity
$v_c$. If the linear density in the streamer is $\lambda\equiv M_s/L$, the
frictional force exerted on it from disc material with impact parameter $b\ll
L$ is
\be 
F_{\rm drag}={4\pi^2 G^2\lambda^2\Sigma_db_1\over v^2_c},
\label{eq:drago}
\ee 
where $b_1$ is the maximum impact parameter considered. This drag force is
exerted only while the streamer intersects the disc. If the streamer follows a
circular orbit of radius $r$, then it intersects the disc for a fraction
$L/(2\pi r)$ of its orbit, so the average drag force is
\be 
\langle F_{\rm drag}\rangle ={2\pi G^2\lambda^2\Sigma_dLb_1\over v_c^2r}=
{2\pi G^2M_s^2\Sigma_d  \over v^2_c r}{b_1\over L};
\label{eq:draga} 
\ee
the maximum impact parameter is $b_1\simeq L$.  Disc material with impact
parameter $b\gg L$ sees the streamer as a point mass, and the time-average of 
the frictional force component from this material is 
\be 
\langle F_{\rm drag}\rangle ={2^{1/2} G^2M_s^2\Sigma_d\ln\Lambda \over
v^2_c r}={2^{1/2} G^2\lambda^2\Sigma_dL^2\ln\Lambda \over v^2_c r},
\label{eq:dragb} 
\ee
where $\Lambda=b_2/b_1$, and $b_2$ and $b_1$ are the maximum and minimum
impact parameters considered. Equations (\ref{eq:draga}) and (\ref{eq:dragb})
are similar except for the Coulomb logarithm in the latter, but the ratio
$b_2/b_1$ is not usually large, so material with $b\ll L$ and $b\gg L$
contributes comparable drag forces. 

Estimating the frictional force is more delicate for long streamers
($\Delta\phi\simgreat 2\pi$). A long streamer intersects the disc several
($N\sim \Delta\phi/2\pi$) times. If the locations of these intersections are
uncorrelated, then the total drag force would be given by the drag per streamer
(eq. \ref{eq:drago} with $b_1\sim r$) times $N$, or
\be 
\langle F_{\rm drag}\rangle ={2\pi G^2M_s^2\Sigma_d \over v^2_c r \Delta\phi}.
\label{eq:dragc} 
\ee
If on the other hand the intersections are localized in the disc--as they
would be if the streamer had a planar orbit of small or moderate
eccentricity--then the $N$ streamers of linear density $\lambda$ act like a
single streamer of density $N\lambda$. Replacing $\lambda$ by $N\lambda$ and
$b_1$ by $r$ in equation (\ref{eq:drago}) we find
\be 
\langle F_{\rm drag} \rangle={G^2M_s^2\Sigma_d\over v^2_cr}.
\label{eq:dragd}
\ee
We regard equation (\ref{eq:dragd}) as more reliable than (\ref{eq:dragc})
in most cases. 

Equations (\ref{eq:draga}) or (\ref{eq:dragb}) for short streamers, and
equation (\ref{eq:dragd}) for long streamers, can be summarized by
\be 
\langle F_{\rm drag} \rangle=g{G^2M_s^2\Sigma_d\over v^2_cr},
\label{eq:drage}
\ee
where $g$ is a dimensionless constant of order unity. The
frictional torque flattens the streamer orbit into the disc plane in a time
\be 
\tau \simeq {M_sv_c\over F_{\rm drag}}\simeq {v_c^3r\over gG^2\Sigma_dM_s}.
\label{eq:str}
\ee 

For a numerical estimate of the flattening time, let $f_s$ be the ratio of the
mass of a streamer to the halo mass inside radius $r$,
$f_s=GM_s/f_hv_c^2r$, and use equation (\ref{eq:isodef}) to eliminate
$\Sigma_d$. Then equation (\ref{eq:str}) yields 
\begin{eqnarray} 
\tau & = & {2\pi\over gf_sf_hf_d}{r\over v_c}\nonumber \\
& = & {3\times
10^{10}\yr\over g}{0.3\over f_h(1-f_h)}
\left({0.01\over f_s}\right)\nonumber \\
& & \qquad \times\left(r\over 3\kpc\right)
\left(200\kms\over v_c\right).
\label{eq:ssss}
\end{eqnarray}
Streamers containing even a few percent of the the halo mass
can be dragged into flattened orbits within a Hubble time.

The principal effect of the frictional torque from the disc is to increase the
$z$-component of the angular momentum of the streamer. The orbital energy of
the streamer changes more slowly: since the gravitational field of the
streamer is approximately stationary, the energy of both the streamer and disc
are conserved. The total angular momentum of the streamer also changes
relatively slowly: for example, the torque on an axisymmetric streamer is
always perpendicular to the streamer plane and hence perpendicular to the
angular momentum vector, so the magnitude of the angular momentum is
unaffected by the friction. This process of increasing $L_z$ at fixed $L$ and
$E$ rotates the streamer planes into the disc plane, without 
strongly affecting their shape. At the same time the disc stars lose $L_z$ at
constant $E$, which heats the disc without strongly affecting the radial mass
distribution. 

To close this section we examine the rate of angular momentum transfer if the
halo is composed of compact objects of mass $M$ (e.g. black holes). Then the
torque on the halo per unit disc area is \cite{O83}
\be 
N=2^{5/2}I_1\ln\Lambda {G^2M\Sigma_d\rho_hr\over\sigma^2}, 
\ee 
where $I_1=0.474$. Using equations (\ref{eq:isodef}) and (\ref{eq:taudef}),
the characteristic flattening time for the halo is then (cf. eq. \ref{eq:gmc})
\begin{eqnarray} 
\tau & = & {\pi \over 2^{3/2}I_1\ln\Lambda f_d}{v_cr^2\over GM}\nonumber \\
& = & 1.4
\times10^{11}\yr\,\left(0.7\over f_d\right)
\left(v_c\over200\kms\right)\left(r\over3\kpc\right)^2\nonumber \\
& & \qquad \times\left(10^6\msun\over
M\right)\left(10\over\ln\Lambda\right). 
\label{eq:bh}
\end{eqnarray}
Observational upper limits to the mass of compact halo objects are
reviewed by \nocite{C94} Carr (1994). Masses $M\simgreat 3\times10^6M_\odot$
are strongly disfavored because they cause excessive heating of the local disc
\cite{LO85,L91}, excessive distortion of gravitationally lensed radio jets
\cite{WP92,G94}, unobserved gravitational lensing of gamma-ray bursts
\cite{Mar98}, and excessive growth of the central black hole in the Galaxy
\cite{xo94}. 

Some authors have argued for much more stringent upper bounds on the mass of
compact halo objects, but we find these arguments unconvincing. (i) Rix \&
Lake (1993) \nocite{RL93} estimate that halo objects with
$M>6\times10^3M_\odot$ would cause excessive heating of the stars in the dwarf
irregular galaxy GR 8. This is a very low-luminosity galaxy ($M_B=-10.6$,
corresponding to $2.5\times10^6L_\odot$ or $2\times10^{-4}L_\star$), so even a
dark halo containing $\sim 30$ times the mass in visible stars will comprise
only $N\sim 10$ halo objects of $3\times10^6M_\odot$. The crossing time in the
halo is $t_c\sim 10^8$ yr, and in $Nt_c\sim 10^9$ yr the halo will evaporate,
leaving a single or binary halo object at the centre, which of course would
not heat the stellar distribution. This argument predicts that the dark mass
in GR8 may be concentrated at its centre, producing a quasi-Keplerian
circular-speed curve, but testing this prediction is difficult: in fact GR8 is
one of the few galaxies with a declining HI rotation curve but there is a
large and uncertain correction for pressure support \cite{CBF90}. (ii) Moore
(1993) \nocite{M93} and Klessen \& Burkert (1996) \nocite{KB96} argue that
massive halo objects disrupt globular clusters; the second of these papers,
which is more thorough and conservative, sets an upper limit
$M<5\times10^4M_\odot$. This result is based on the assumption that the
disruption time $t_d$ for a particular subset of globular clusters (central
density between $10^3$ and $10^4M_\odot\hbox{ pc}^{-3}$) must be long enough
that more than 1\% of them survive; for a Poisson process with destruction
probability per unit time $t_d^{-1}$ this requires that $t_d>0.22t_H$ where
$t_H$ is the age of the cluster population. However, if the present clusters
represent the survivors of a much larger original population--as indicated by
the efficiency of the destruction mechanisms associated with known components
of the Galaxy \cite{GO97,MW97}--then we expect a wide range of survival
times in any snapshot of the population. In particular, if the distribution of
lifetimes at formation is a power law, then we expect the fraction of clusters
with lifetimes less than $t_d\ll t_H$ to be $\sim t_d/t_H$ so $\sim 20\%$ of
clusters would be {\it expected} to violate the Klessen-Burkert criterion.

We conclude from equation (\ref{eq:bh}) that if the halo is composed of
$10^{6}\msun$ black holes, well below the allowed upper limit of
$3\times10^6M_\odot$, then we expect substantial halo flattening and rotation
out to $\sim 1$ kpc, which may be sufficient to suppress halo dynamical
friction on rapidly rotating bars.

\section{Constraints on halo shapes}

The arguments in the preceding sections suggest that the inner parts of dark
halos may be thick, rapidly rotating discs. As noted earlier, bars are
strongly coupled to the inner halo and thus should also rotate rapidly, as
observed. We now ask whether other observations constrain the shapes of inner
halos. Although halos formed by hierarchical clustering are generally
triaxial, the subsequent growth of the disc tends to make the inner halo more
nearly axisymmetric (cf. \S 1), and this expectation is confirmed by the
upper limits to the ellipticities of most disc galaxies induced by the
halo potential--typically $\simless 0.1$; see \cite{FZ92,KT94,R96}. Thus we
assume that the inner halo is axisymmetric, and restrict our attention to the
flattening of axisymmetric halos.

Rix (1996)\nocite{R96} and Sackett (1996, 1998)\nocite{S96,Sa98} review the 
tools that have been used to constrain the halo shape in our own and other
galaxies, including the shapes of X-ray halos, the kinematics of polar-ring
galaxies, the thickness of HI layers, and gravitational lensing.

The kinematics and spatial distribution of Population II objects can be used
to constrain the shape and mass of the dark halo in our Galaxy
\cite{MRS81}. Unfortunately the best available results are model-dependent
and have large error bars: $0.3 < c/a < 0.6$ from Binney, May \& Ostriker
(1987) \nocite{BMO87} and $c/a>0.34$ from \nocite{vdM91} van der Marel (1991).

The ellipticity of the mass distribution in edge-on disc galaxies can be
determined from the shape of their X-ray isophotes. This method yields
$c/a=0.60\pm0.13$ for the S0 galaxy NGC 1332 \nocite{BC96} (Buote \&
Canizares 1996); this is greater than expected for disc-like dark matter, but
the uncertainties are large and the result is dominated by halo radii well
beyond the outer edge of the disc, where the flattening mechanism discussed
here would not operate.

Steiman-Cameron, Kormendy \& Durisen (1992) \nocite{SC92} fit a
precessing-disc model to the dust lanes in the S0 galaxy NGC 4753 and derive
$0.84< c/a < 0.9$; however, this result depends on an assumed age for the
event that warped the galaxy disc.

The axis ratios of the dark halos in polar-ring galaxies have been estimated
by several investigators. Whitmore, McElroy \& Schweizer \nocite{W87} (1987)
examined three S0 galaxies with polar rings and found axis ratios for the
potentials of 0.86 to 1.05 ($\pm0.2$), corresponding to axis ratios for the
density of roughly 0.58 to 1.15 ($\pm 0.6$); however, two of these galaxies
were re-examined by \nocite{S94,SP95} Sackett et al. (1994) and Sackett \&
Pogge (1995), who found much flatter halos with axis ratios 0.3--0.5.

High-resolution HI data can be used to determine the thickness of the HI layer
in highly inclined disc galaxies; at and beyond the edge of the optical disc
this thickness is sensitive to the shape of the dark halo. The most thorough
analysis of this kind, for the Scd galaxy NGC 4244 \cite{O96}, yields
$c/a\simeq0.2{+0.3\atop -0.1}$. Olling \& Merrifield (1997)\nocite{OM97}
derive $c/a=0.75\pm0.25$ for the Galaxy using both the thickness of the HI
layer and local stellar kinematics. 

Gravitational lensing by disc galaxies can constrain the halo mass and
shape. Koopmans, de Bruyn \& Jackson \nocite{KBJ98} (1998) find $c/a>0.5$ for
the edge-on lensing galaxy B1600+434. Flattening the halo does not
significantly affect the overall lensing cross-section of a galaxy but does
enhance the fraction of images that exhibit characteristic geometries
associated with disc lensing \cite{KK98}. Unfortunately, disc galaxies are
expected to comprise only 10--20\% of gravitational lenses so a large sample
will be difficult to obtain.

Evidently the present data allow a wide range of halo shapes, with some
indication that moderately flattened halos ($c/a\sim 0.5$) are more common
than spherical or disc-shaped halos. The data provide no strong evidence for
or against rotating thickened discs of dark matter such as those proposed
in this paper. 

\section{Discussion}

Structure in galactic discs and dark halos couples these two components
together, leading to angular momentum transfer from disc to halo, spinup and
flattening of the inner halo, and heating and thickening of the disc. The
time-scale for this process is highly uncertain. Known structures in the disc
are probably not able to flatten and spin up the inner halo significantly in a
Hubble time, but halo structure is likely to be more efficient. In particular,
we have identified two types of halo structure that may be able to spin up the
halo: massive black holes, and tidal streamers resulting from hierarchical halo
formation.

A rotating flattened halo may be needed to explain the high pattern speed of
bars (\S \ref{sec:intro}). Flattening the inner halo also thickens and heats
the inner disc; this process may form some galactic bulges, which, like discs,
are metal-rich and rapidly rotating \cite{GW97}. Another way to make bulges
from discs is through a buckling instability of bar structures in the discs
\cite{CS81,RSJK91}. The hypothesis that some bulges may be produced from
discs is supported by the inner cutoffs observed in disc surface-brightness
profiles as they enter the bulge \cite{K77}, and the similar scale lengths
of inner discs and bulges \cite{C96}. Kormendy (1993) \nocite{K93} has
argued persuasively that some bulges are really discs in terms of their
dynamics and origin.  On the other hand, the centres of bulges have much
higher phase-space density than discs and so cannot arise from discs through a
collisionless process \cite{C86}; moreover the strong similarity between
bulges and elliptical galaxies of similar luminosity suggests that most bulges
are formed in a manner similar to the family of equivalent ellipticals.

Dynamical friction on tidal streamers provides a natural mechanism to produce
flattened dark halos of non-baryonic material, and thus offers a
counter-example to the usual belief that if the dark halo is disclike it must
be baryonic.  If this process is important then we expect the inner halos of
spiral galaxies to be substantially flatter than halos of ellipticals. 

Relaxation from tidal streamers can also be important in elliptical
galaxies. Gravitational scattering of stars by streamers can enhance the
relaxation rate compared to the usual Chandrasekhar formulae for two-body
relaxation \cite{BT87} and hence isotropize the distribution function
at radii where two-body relaxation is ineffective (energy relaxation is less
effective than isotropization because the potential from a long streamer only
varies slowly in time).

We may contrast relaxation from tidal streamers with two other relaxation
processes in stellar systems: two-body and violent relaxation. The two-body
relaxation rate is determined by assuming that the stellar system is as smooth
as possible, so the only potential fluctuations arise from Poisson noise due
to individual stars. Violent relaxation \cite{LB67} is modelled by assuming
that the potential fluctuations in the stellar system are as large and rapid
as possible (spatial and time scales of order the system size and dynamical
time), an assumption that is only valid during the initial collapse of the
system. Relaxation from tidal streamers is intermediate--stronger than
two-body relaxation, and weaker but longer lasting than violent
relaxation--and arises because a stellar system only phase mixes gradually
over many dynamical times. Perhaps the process should be called `non-violent
relaxation'. 

One important unresolved question is how and when the streamers are finally
mixed together. Small-scale irregularities in the disc and halo (\gmcs,
globular clusters, etc.) can mix the streamers, but relaxation arising from
interactions between streamers may be more effective. This issue is also
important for experiments that hope to detect the phase-space structure of
dark matter particles \cite{SI92}. 

The proposals made in this paper can be tested in a variety of ways. The
physical processes can be studied by appropriate numerical simulations.
Conventional $N$-body simulations of galaxy formation offer limited insight
into phase mixing, because of numerical noise
\cite{HB90,HO92,SW97}. However, specially designed simulations are more
powerful. A reasonable approximation is to turn off the gravitational forces
among the disc particles and among the streamer particles, to suppress
large-scale instabilities in the disc and numerical relaxation within the
halo. The only remaining forces would be between disc and streamer
particles, and between both sets of particles and a fixed galactic potential.
A more complete treatment would include the mutual gravitational
interactions between streamers. As an illustration of the importance of such
interactions, consider a single streamer consisting of an axisymmetric ring of
material in an inclined orbit. Eventually this streamer would be dragged
precisely into the disc midplane. However, if there are several axisymmetric
streamers with the same radius and different midplanes, they cannot {\em all}
be dragged into the disc midplane since this would lead to an increase in the
phase-space density of the halo material, which violates the collisionless
Boltzmann equation.

There are also direct observational tests. (i) Massive black holes scatter a
fraction of the stars in the bulge and inner disc out to much larger radii;
these should appear as a distinct population of $\sim 10^8M_\odot$ in
metal-rich high-velocity stars on nearly radial orbits (tidal streamers are
less effective scatterers because their potential is softer). Similarly, black
holes scatter local disc stars to large epicyclic energies, creating a
power-law tail containing $\sim 1\%$ of the disc mass \cite{LO85}. (ii) Tidal
streamers heat disc stars by removing angular momentum at constant energy,
while massive black holes induce a random walk in both energy and angular
momentum; these processes may leave distinct signatures in the phase-space
distribution of inner disc and bulge stars. (iii) Heating by tidal streamers
is most effective early in the galaxy's history, while heating by black holes
continues throughout its lifetime; this distinction may be reflected in the
age or metallicity distribution of bulge stars. (iv) The presence of a thin
disc of old stars near the centre of a galaxy would imply that substantial
angular-momentum transfer from disc to halo has not occurred since the disc
was formed. Some early-type galaxies contain discs with scale lengths
$\simless 1 \kpc$ \cite{BJM98,SBWCM98}; if these are old and edge-on they can
be used to constrain the disc thickening and hence angular-momentum transfer
rates. (v) Halo structure may be directly detectable from small-scale
fluctuations in the surface-brightness distribution of elliptical galaxies;
these are distinct from surface-brightness fluctuations arising from bright
stars and globular clusters (e.g. \nocite{JTL98} Jensen, Tonry \& Luppino 1998)
because they are correlated in nearby pixels. Halo structure may also account
for discrepancies in the radio flux ratios predicted by smooth mass models of
gravitational lensing \cite{MS98}.

\section*{Acknowledgments}

We thank Jerry Sellwood for informative discussions.  This research was
supported in part by NASA grant NAG5-7066 and NSF grant AST-9424416.

\label{lastpage}

\end{document}